# Towards Reverse-Engineering the Brain:
# Brain-Derived Neuromorphic Computing Approach with Photonic, Electronic, and Ionic Dynamicity in 3D integrated circuits

S. J. Ben Yoo[1], Luis El-Srouji[1], Suman Datta[2], Shimeng Yu[2], Jean Anne Incorvia[3], Alberto Salleo[4], Volker Sorger[5], Juejun Hu[6], Lionel C Kimerling[6], Kristofer Bouchard[7], Joy Geng[1], Rishidev Chaudhuri[1], Charan Ranganath[1], Randall C. O'Reilly[1]

## A. ABSTRACT:

The human brain has immense learning capabilities at extreme energy efficiencies and scale that no artificial system has been able to match. For decades, **reverse engineering the brain** has been one of the top priorities of science and technology research. Despite numerous efforts, conventional electronics-based methods have failed to match the scalability, energy efficiency, and self-supervised learning capabilities of the human brain. On the other hand, very recent progress in the development of new generations of photonic and electronic memristive materials, device technologies, and 3D electronic-photonic integrated circuits (3D EPIC [1]) promise to realize new **brain-derived** neuromorphic systems with comparable connectivity, density, energy-efficiency, and scalability. When combined with bio-realistic learning algorithms and architectures, it may be possible to realize an "**artificial brain**" prototype with general self-learning capabilities. This paper argues the possibility of reverse-engineering the brain through architecting a prototype of a *brain-derived* neuromorphic computing system consisting of artificial electronic, ionic, photonic materials, devices, and circuits with dynamicity resembling the bio-plausible molecular, neuro/synaptic, neuro-circuit, and multi-structural hierarchical macro-circuits of the brain based on well-tested computational models (Leabra simulator). We further argue the importance of bio-plausible local learning algorithms applicable to the neuromorphic computing system that capture the flexible and adaptive unsupervised and self-supervised learning mechanisms central to human intelligence. Most importantly, we emphasize that the unique capabilities in brain-derived neuromorphic computing prototype systems will enable us to understand links between specific neuronal and network-level properties with system-level functioning and behavior. Among the many possible questions that can be addressed, three specific examples include: (1) <u>Learning processes in sparse hierarchical networks;</u> (2) <u>Oscillations and long-range communication and coordination;</u> and (3) <u>Brain Network Efficiency</u>. We believe that these studies can reveal how the brain is capable of accomplishing massive computational tasks with high energy efficiency.

1. University of California, Davis, CA, USA
2. Georgia Institute of Technology, Atlanta, GA, USA
3. University of Texas, Austin, TX, USA
4. Stanford University, Stanford, CA, USA
5. University of Florida, Gainsville, FL, USA
6. Massachusetts Institute of Technology, MA, USA
7. University of California, Berkeley, CA, USA

## B. Introduction

The human brain has immense learning capabilities at extreme energy efficiencies and scales that no artificial system has been able to match. The human brain recognizes or associates features from partial and conflicting information at ~20 W power levels. The brain is bombarded with vast amounts of sensory information, but somehow, it makes sense of this data stream, even if it contains imperfect and inconsistent data elements; it does this by extracting the spatiotemporal structure embedded in the data stream and builds meaningful representations through parallel distributed processing. In humans, each neuron may be connected to up to ~10,000 other neurons, passing signals via as many as 1,000 trillion synaptic connections, equivalent by some estimates to a computer with a 1 trillion bit per second processor at ~10 MW power level.

For decades, *reverse engineering the brain* has been one of the top priorities of science and technology research. Despite numerous efforts (e.g., IBM's TrueNorth and Northpole [2][3], Intel's Loihi [4][5], University of Manchester's SPINNAKER [6]), conventional electronics-based methods have failed to match the scalability, energy efficiency, and self-supervised learning capabilities of the human brain. Current approaches in Artificial Intelligence (AI) are based on computational architectures that are loosely inspired by the brain. While the AI agents have been able to outperform humans at complex games like "Go" and "Chess" when trained on a relatively narrow range of tasks, they often require a considerable degree of training, and they are brittle in the face of unexpected changes to surface-task characteristics. Recently developed Generative Pre-trained Transformer (GPT) [7]–[9] show apparent ingenuity, however, requires pretraining (no online learning) with more than 12 million USD worth of energy on GPU-based systems just for training for GPT-3 [10] and far more for GPT-4[7], [9]. The human brain, in contrast, is capable of remarkably fast learning in a manner that is flexible and enables generalization to new situations and tasks, and it does so with a remarkably low level of energy consumption relative to traditional computational hardware. Here, we opine that a **brain-derived**—rather than a *brain-inspired*—architecture will lead to a paradigm shift, enabling the development of intelligent agents that can work in tandem with humans on complex tasks in noisy, unpredictable environments. Key to this paradigm shift will be **the integration of hardware** that <u>emulates the brain's processing characteristics with design principles based on fundamental aspects of neural plasticity and circuit design in the human brain</u>.

Very recent progress in the development of new generations of photonic and electronic memristive materials, device technologies, and 3D electronic-photonic integrated circuits (3D EPIC [1]) promise to realize new **brain-derived** neuromorphic systems with connectivity, density, energy efficiency, and scalability comparable to that of the biology. When combined with bio-realistic learning algorithms and architectures, it may be possible to realize an "**artificial brain**" prototype with general self-learning capabilities. This paper argues the possibility of reverse-engineering the brain through architecting a prototype of a *brain-derived* neuromorphic computing system consisting of artificial electronic, ionic, photonic materials, devices, and circuits with dynamicity resembling the bio-plausible molecular, neuro/synaptic, neuro-circuit, and multi-structural hierarchical macro-circuit of the brain. We further argue the importance of bio-plausible local learning algorithms applicable to the neuromorphic computing system that capture the flexible and adaptive unsupervised and self-supervised learning mechanisms that are central to human intelligence.

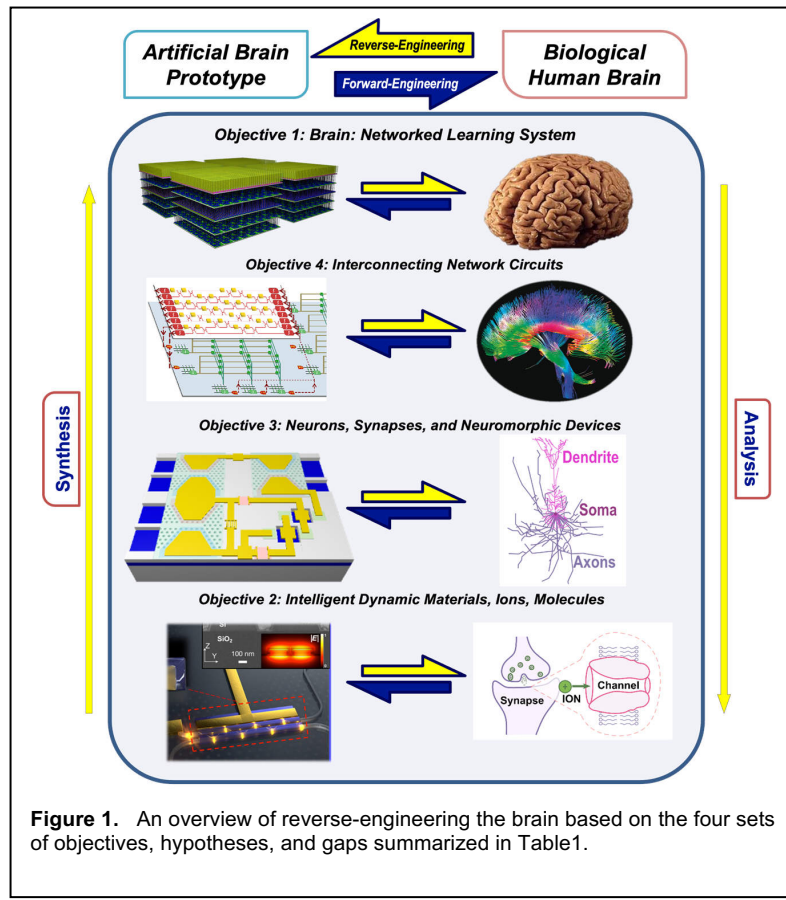

**Figure 1.** An overview of reverse-engineering the brain based on the four sets of objectives, hypotheses, and gaps summarized in Table1.

To address the failures of previous efforts in reverse-engineering the brain, we identify four fundamental scientific and technological "Gaps" summarized in **Table 1**. By exploiting advances in electronic, ionic, and photonic technologies at the molecular, device, circuit, and system levels to emulate the brain (**Figure 1**), the newly developed advanced technologies and neuroscience concepts can possibly address these gaps to pursue reverse-engineering the brain through brain-derived approaches under specifically relevant hypotheses. If successful, the resulting neuromorphic computing paradigm may reproduce the uniquely flexible and adaptive nature of human intelligence with extreme scalability and energy efficiency. The effort to reverse-engineer the brain will also advance our understanding of distributed learning in the brain and neuromorphic systems through a synergistic research effort. Novel neuroscientific insights will inform advanced neuromorphic hardware, and that hardware will, in turn, help to understand how interactions across brain systems enable the uniquely flexible and adaptive nature of human intelligence.

**Table 1.** Possible four 'Gaps' commonly seen in the previous neuromorphic computing research activities, and the corresponding *Hypotheses* and the *Objectives* of the Brain-derived methods for neuromorphic computing to reverse-engineer the brain.

| |
|---|
| ***Gap 1: Lack of understanding of the principles of learning, plasticity, and dynamics in the context of networked neurons with structured connectivity*** <br> **_Hypothesis 1:_** Human-level intelligence emerges through dynamic networked interactions between multiple specialized brain systems with structured and efficient large-scale hardware connectivity. <br> **_Objective 1:_** Develop a comprehensive simulator and build a novel neuromorphic computing prototype system that incorporates insights from cutting-edge modeling and experiments about synaptic plasticity, network dynamics, and learning in cortical circuits, and fundamental attributes of human learning and memory. In reverse, utilize the prototype system to understand the brain. |
| ***Gap 2: Lack of methods to realize neuromorphic dynamics in bio-derived materials*** <br> **_Hypothesis 2:_** Conventional electronic materials such as silicon are unable to faithfully replicate the dynamicity driven by ions, molecules, and structural changes in the dendrites, synapses, and somas. <br> **_Objective 2:_** Pursue new photonic, electronic, and ionic memristive materials that can closely resemble the dynamic mechanisms responsible in the biological neural systems. |
| ***Gap 3: Lack of methods to realize brain-derived neuromorphic devices*** <br> **_Hypothesis 3:_** Conventional electronic devices such as CMOS transistors and electrical wires are unable to faithfully replicate the dynamicity seen in the dendrites, synapses, and somas. <br> **_Objective 3:_** Pursue new photonic, electronic, and ionic memristive devices that can closely resemble the dynamic mechanisms seen in the biological neural systems. |
| ***Gap 4: Lack of scalable and energy-efficient interconnecting circuits for brain-like hierarchical learning*** <br> **_Hypothesis 4:_** Current (analog) electronic approaches are unable to achieve the connectivity (e.g. ~8000 synaptic connections per neuron) at scale (e.g. billions of neurons) limited by electronic wirings. <br> **_Objective 4:_** Pursue 3D photonic-electronic integrated circuits that offer high density and high connectivity with extreme efficiency at scale while supporting hierarchical learning in optical macro-circuits and electronic micro-circuits. We will conduct simulation and experimental testbed studies. |

**Figure 2** expands the view of **Figure 1** to emphasize brain-derived biophysical principles and algorithms applied in a hierarchical network [11] of spiking optoelectronic neural networks (macro-circuits) and spiking electronic neural networks (micro-circuits), including nanoscale dendrite computing (nano-circuits). Combined with bio-inspired intelligent ionic electronic and photonic materials in the neural networks constructed in 3D, we expect the resulting neuromorphic system to realize ***brain-derived complex computing elements and learning mechanisms that span multiple timescales*** on a new platform that combines ***low-noise, scalability, wavelength-parallelism, high-throughput, and dynamic memristive plasticity*** [12] of **photonics**, and ***high-density, agility, and dynamic plasticity*** [13], [14] of **electronics**.

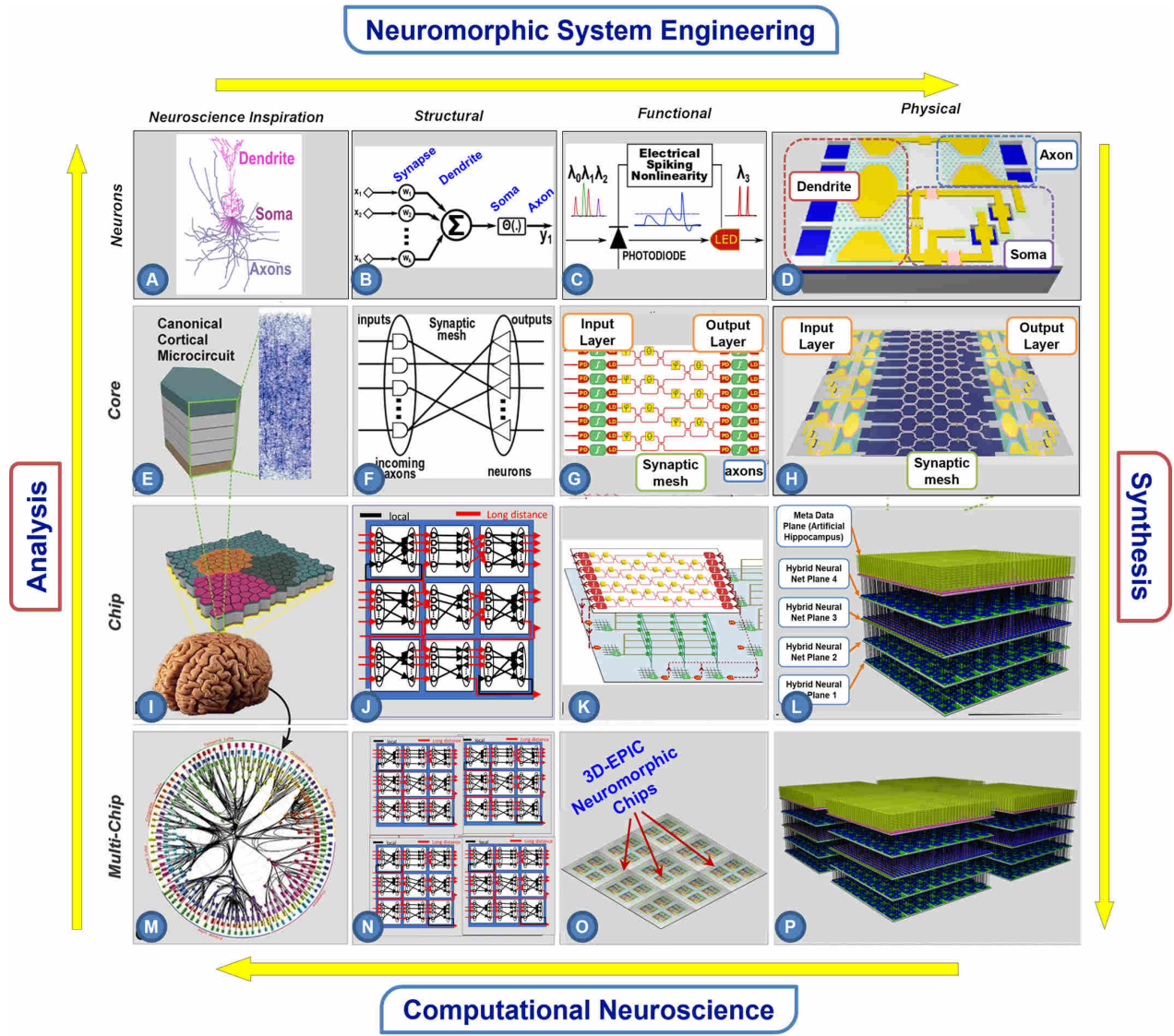

**Figure 2.** A conceptual 3D Hierarchical Neuromorphic Nanocomputing architecture extending on the framework by [15]. **(A)** A canonical neuron, **(B)** Neuron's minimal structure, **(C)** Neuron's simplified functional diagram (optoelectronic neuron example), **(D)** A physical schematic for a nanoscale optoelectronic neuron, **(E)** cortical microcircuit, **(F)** Structure of a neurosynaptic core with axons as signal carriers (inputs/output), synapses as directed connection strength, and neurons as nonlinearity. **(G)** Functional view of a photonic synaptic mesh between presynaptic and post synaptic neurons. **(H)** Physical layout of (G). **(I)** A two-dimensional map of cortical columns in a functional network. Multichip scales are both created by interconnecting **(J)** neuron microcircuits reconfigurable optical synaptic interconnects. **(K)** Hybrid optical (red) and electronic (green) neural network forming a hierarchical macrocircuit. **(L)** Schematic of 3D electronic photonic integrated circuit (EPIC) neural network consisting of multiple planes of (K). **(M)** Illustration of long-range connections between cortical regions in the macaque brain [16]. **(N)** Interconnections of many functionally specialized neural macro-circuits (J). **(O)** Multi-3D EPIC chip neural networks emulating functional specializations of interconnected human brain structures. **(P)** Schematic of (O).

### B.1. Nano-electronic Brain-derived Computing

*Background*: Spatiotemporal activity patterns in the brain are the basis for higher-order cognitive functions such as real-time learning. Formal approaches (theoretical and empirical) to studying neural collective phenomena and associating specific patterns of spatiotemporal brain activity to continual learning and real-time adaptation are necessary and markedly different from current approaches for brain-inspired neuromorphic hardware. Today's brain-inspired hardware (a) implements models that are low-dimensional versions of the neural structures of interest, (b) produces a single behavior that is hardwired into the system, and, hence, c) lacks adaptation and flexibility (Gap 1-4 of **Table 1**).

***Objectives*:** We should take new approaches to pursue hierarchical implementation of continuous-time dynamical systems (CTDS) using novel electronic materials, devices, and coupling elements and to find new ways to prime and place the system near the criticality point of seamless transitions. The resultant hardware will possess the capacity to adapt and respond to complex patterns of agile data and infer in real-time to take appropriate action. Driven by Hypotheses 1-4 (**Table 1**), we should investigate the individual and collective dynamics of nano-electronic devices and their coupled networks, pursue new genres of brain-derived nano-materials/devices, exploit advances in monolithic 3D nano-electronic integration to enable dense and reconfigurable connectivity, and interface with photonic networks to implement brain-derived scale-invariant neural hardware with a range of emergent dynamics poised near the criticality point.

This paradigm, the collective approach to neural computation, departs from the traditional connectionist approach in which the interactions between logic and memory are programmed using a central controller. Hence, this paradigm of nano-electronic brain-derived computing for reverse-engineering the brain can be broken into the following two goals:

- Synergistically explore the implementation of a vast repertoire of spatio-temporal activity patterns following the networked learning principle at different levels of the hierarchy (materials, devices, networks, and computational theory).
- Design a hierarchical and reconfigurable network of interacting nano-electronic devices and circuits to build a CTDS in which the properties of the neural computation task (learning and adaptation) are encoded within the dynamics of the material or the network itself so that the resulting computation in the CTDS can span vast spatial and temporal scales.

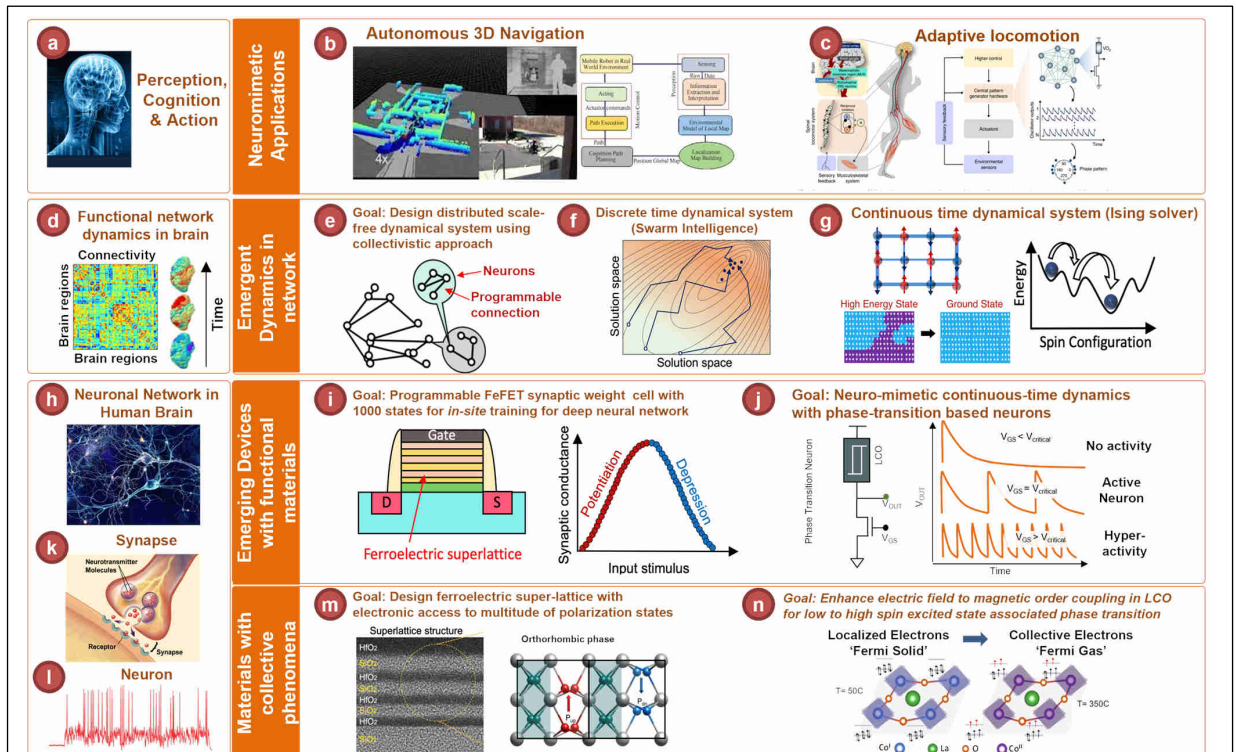

**FIGURE 3.** Brain-derived Nano-electronic Computing framework with intelligent materials, biomimetic devices and reconfigurable circuits. (a-c) Examples of applications for brain-derived functional blocks mimicking cognitive structures in the human brain. (d-g) Examples of emergent spatio-temporal patterns exhibited by artificial dynamical systems that resemble human brain activity. (h-j) Emerging devices replicating functional characteristics of biological neurons. (k-n) Study of collective phenomena emerging from symmetry breaking in intelligently engineered materials.

### *B.1.1. Neural Components: Neurons, Synapses and Connectivity Network*

**Fig. 3** illustrates how the collective networked neuromorphic computing approach, a fundamental departure from the traditional connectionist approach, can be achieved concurrently and synergistically at all levels of the compute hierarchy towards realizing a **truly scale-free distributed compute capability**. Heterogeneous and diverse electronic synapses and neurons can be enabled by diverse intelligent nano-electronic materials ranging from ferroelectric and ferromagnetic materials to ionic-graphene and electro-chemical materials. The variety of materials available—polymers, MXenes, and graphene— enable a great span of device properties in terms of dynamic range of tunable conductances, switching time (down to ns) and switching energies (as low as 50 aJ/µm² [17]).

Dendrite computing is a must for realizing high-density and high-energy-efficiency neuromorphic computing. Dendrites branch out from the neuronal body and process spike information into spatiotemporal signals. For example, the leaky integrate-and-fire model of neuronal behavior is composed of one leaky recurrent dendrite plus a soma for an activation. In many biological systems, each neuron has multiple sets of dendrites independently processing incoming spikes, allowing for a wide variety of neuronal configurations [18]. Artificial dendrite function is an underexplored topic in neuromorphic computing. Some work has been done in this area, for example, analog neuromorphic devices for artificial dendrites using CMOS circuitry to emulate dendritic functions [19]–[22]. Another work experimentally demonstrated a dendrite-like device [23] realized on a multi-gate ferroelectric FET device to emulate selective spatio-temporal pulse sequence processes in dendrites with 100× signal-margin. More recent work utilized dual-gate operation of the graphene devices, and demonstrated that the devices emulate three different dendritic spatiotemporal signals: leaky recurrent, alpha, and gaussian [24]. The devices can work independently or combined to create higher complexity time-dependent dynamics.

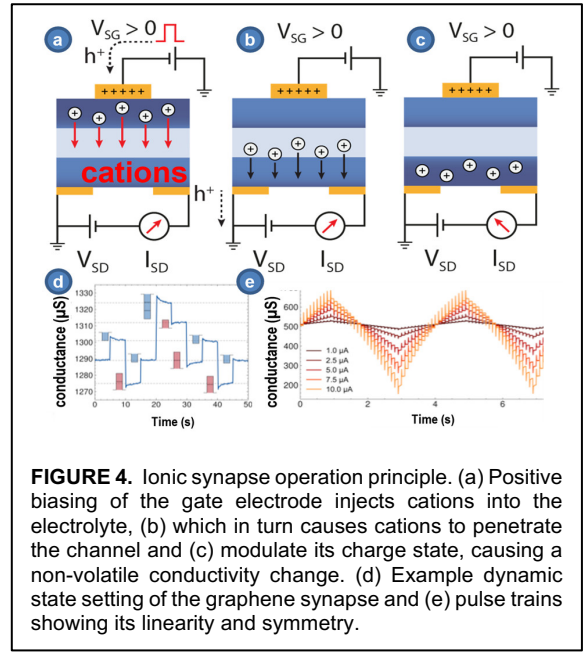

**FIGURE 4.** Ionic synapse operation principle. (a) Positive biasing of the gate electrode injects cations into the electrolyte, (b) which in turn causes cations to penetrate the channel and (c) modulate its charge state, causing a non-volatile conductivity change. (d) Example dynamic state setting of the graphene synapse and (e) pulse trains showing its linearity and symmetry.

Electrochemical artificial synapses, as shown in **Fig. 4,** can utilize two semiconductor films separated by an electrolyte [13], wherein one acts as gate electrode and controls the conductance state of the other in analog fashion. Device operation is based on reversible ion insertion/extraction from the electrolyte into the semiconductor channel driven by the gate electrode potential. These ionic artificial synapses exhibit fast switching (20ns write), sub-µs write-read cycling, low voltage (±1V) and low current (µA to nA) operation [25], [26] They have a high density of non-volatile states (>6bit) that can be linearly and symmetrically programmed with a high signal-to-noise ratio of ~ 100, while having excellent endurance of >$10^9$ read-write events. The analogy between neurotransmitters diffusing through the synaptic cleft and ions drifting in and out of the channel of the artificial synapse is apt as these devices show both long-term potentiation and short-term potentiation depending on the device architecture (i.e. polarizable vs. non-polarizable gate electrode [13], [27]) and metaplasticity (i.e. potentiation depending on device state), leading to demonstration of spike timing dependent plasticity (STDP) for Hebbian learning. In addition, artificial synapses and neurons based on monolayer graphene transistors can achieve similar dynamical responses as their biological counterparts (~50 aJ/µm2 switching energy at ~ms switching time). The heterogeneity of these materials and devices in the neural networks offer diversity in dynamicity, which also help create bio-plausible systems where Hebbian learning processes can strengthen or weaken synaptic connections with preferred or non-preferred neurons. Graphene is also promising for electrical interconnects capable of emulating dynamically reconfigurable interconnect networks to implement dendritic activity. Here, operating artificial synapses closer to the extremes of their transfer-functions will allow a non-linear response and saturation, which can be used as a local activation function. Wong recently demonstrated an *inference* chip [28] with resistive switching random-access memory (RRAM) synapses monolithically integrated with foundry-fabricated CMOS neurons and control circuits to perform a variety of AI tasks. The *learning* process in addition to inference can leverage the ability to modulate the channel conductance offered by artificial synapses. Recently, back-end-of-line (BEOL) compatible wide band gap semiconducting oxide transistors have been reported with ultra-low off-state leakage current < 10 fA/µm [29]–[32], more than 10,000 times lower than that of silicon transistors, enabling learning and inference without relying on external non-volatile memories.

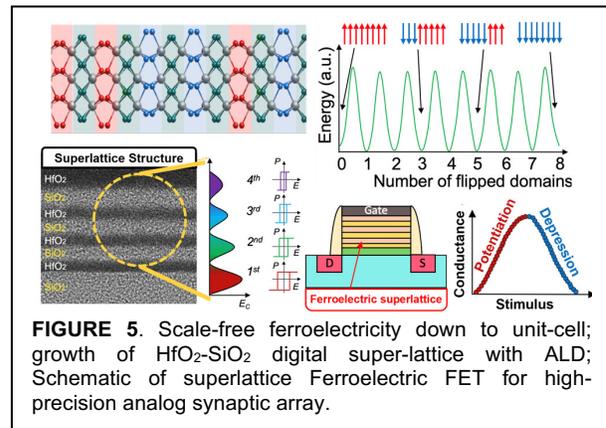

**FIGURE 5**. Scale-free ferroelectricity down to unit-cell; growth of $HfO_2$-$SiO_2$ digital super-lattice with ALD; Schematic of superlattice Ferroelectric FET for high-precision analog synaptic array.

### B.1.2. Reconfigurable Neural Network with Emergent Dynamics:

Ferroelectricity arises from *breaking spatial inversion symmetry* resulting in spontaneous ordering of electric dipoles in a crystal. Hafnium dioxide ($HfO_2$) has been shown to exhibit ferroelectricity and is fundamentally different from traditional ferroelectric with $ABO_3$ perovskite structure. Datta and Salahuddin have experimentally demonstrated dipole-by-dipole switching of polarization in zirconium-doped $HfO_2$ (HZO) using amplitude-modulated electrical pulses for analog synapse application for *in situ* learning of 8-bit precision weights in artificial neural networks. First-principles calculations show that flat bands of polar phonons and localized dipoles induce a scale-free ferroelectric order in $HfO_2$. Negligible interactions between each vertical dipole in the ferroelectric HZO thin film results in a scale-free behavior of unit-cell-wide ferroelectric and offers the extra-ordinary opportunity to construct high-precision (>1,024 levels) analog synaptic memory by exploiting its unit-cell level switchability (**Fig. 5**). High-precision analog synapses whose conductance characteristics exhibit linear and symmetric response to input stimuli allow the network to learn in real-time and memorize the weights as part of real-time adaptation and life-long learning [33], [34]. For scale-free learning, Datta has experimentally demonstrated an Ising machine [34](**Fig. 3n**) consisting of a network of electrically coupled phase transition nano-oscillators (PTNOs) driven by a second harmonic injection locking signal. Here, time-symmetry breaking stabilizes a repertoire of many-body states in a coupled network of parametrically driven nano-oscillators. A nonlinear dissipative oscillator of natural frequency ($\omega$) displays period doubling when it is modulated at twice the frequency $2\omega$. The emerging period-2 states have opposite phases and can be associated with up and down spin states. The symmetry of the effective Hamiltonian for any oscillator can be broken in the presence of another and the switching rates between the two minima diverge. Depending on the nature of coupling, the "deeper" well corresponds to the oscillators having the same (ferromagnetic) or the opposite (anti ferromagnetic) phase [35]. The abrupt insulator-to-metal phase transition in a correlated oxide system (lanthanum cobalt oxide, LCO) can be utilized to realize a compact nano-oscillator. The phenomenon of synchronization and emergent spatiotemporal pattern from the interaction among coupled oscillators can be exploited to generate a range of locomotion gait patterns. The coupled oscillators exhibit stable limit-cycle oscillations and tunable natural frequencies for real-time programmability of phase-pattern. This can lead us to implement a dynamical system using coupled phase transition nano-oscillators and analog ferro FETs as bidirectional coupling elements that stabilize the criticality point of a multitude of spatio-temporal phase patterns and prime the system to learn, adapt, and respond in real-time to external stimuli. It will enable artificial agents to navigate a complex and critical world, and to respond or adapt to unexpected, surprising events in real-time.

### B.2. Nano-optoelectronic Neuromorphic Computing

***Background:*** The repertoire of spatiotemporal activity patterns in the brain is the basis for higher-order cognitive functions like real-time learning and adaptive response (Gap 1-4 of **Table 1**). However, today's brain-inspired hardware based on electronics has limited scalability, interconnectivity, and energy efficiency. Photonic information processing and interconnects with high-level signal multiplexing in wavelength division (WDM) to overcome these challenges. Recent success in nanoscale optoelectronic (OE) neurons, nanophotonic synaptic neural networks, and photonic Ising machines [36] demonstrate potential for brain-derived nano-optoelectronic neuromorphic prototypes.

***Objectives:*** We should investigate nano-optoelectronic neuromorphic computing features numerous examples of innovations toward realizing OE circuits with the emergent properties of a human brain.

- Bio-derived artificial OE neurons [37]–[41] realized by co-designed nanophotonic and nano-electronic integrated circuits exhibit dynamic characteristics of biological neurons.
- Nanophotonic synapses incorporating a new generation of photonic memristive materials and OE devices exhibit bio-plausible dynamicity and plasticity [42]–[45] of biological systems.
- Photonic neural network realizes wavelength and spatial diversity [46]–[50] with photonic tensor-core

decomposition [51], [52] aiming at achieving scalability of the human brain.
- The simulator developed here will benchmark performance, energy-efficiency, and throughput [41], [53]–[56] of the OE neural network, will help create 3D EPIC simulator, and will help investigate the normative behavior of the multi 3D EPIC artificial brain prototype.

### B.2.1. Photonic Memristive Materials & Devices

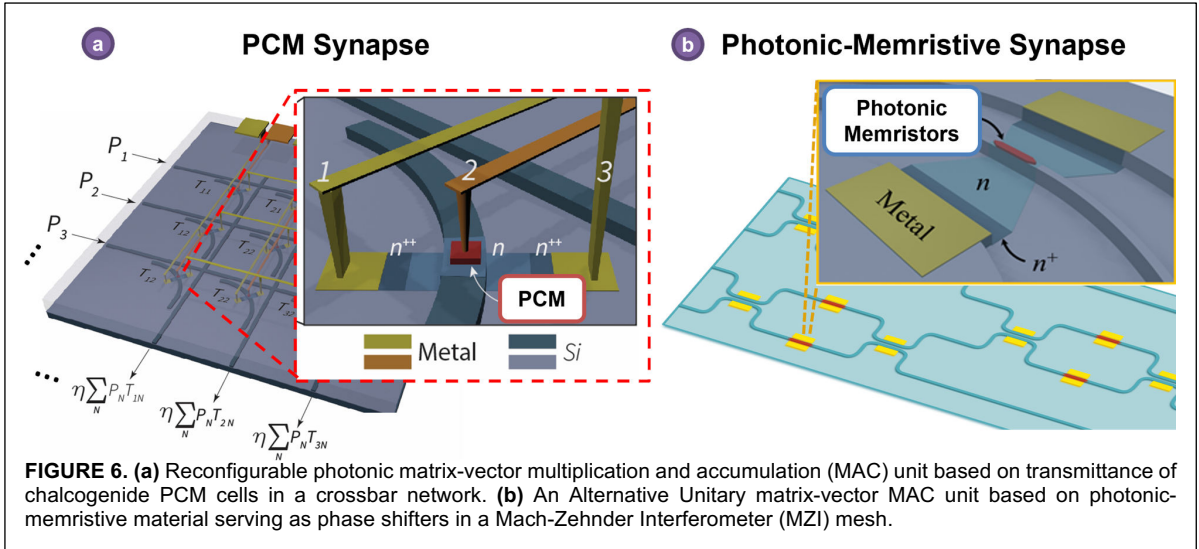

**FIGURE 6. (a)** Reconfigurable photonic matrix-vector multiplication and accumulation (MAC) unit based on transmittance of chalcogenide PCM cells in a crossbar network. **(b)** An Alternative Unitary matrix-vector MAC unit based on photonic-memristive material serving as phase shifters in a Mach-Zehnder Interferometer (MZI) mesh.

Dynamic photonic materials capable of self-reconfiguration are necessary for realizing plasticity in photonic synapses in neural networks. Chalcogenide glasses represent a class of dynamic materials with reconfigurable optical, electrical, and ionic properties emulating the biological synaptic plasticity. The materials can be switched either by light or electrical pulses to access a continuum of states between its amorphous and crystalline phases while featuring a large tuning range in both optical refractive index ($\Delta n$ ~ 1) and electrical conductivity (4 to 5 orders of magnitude). Another key attribute of chalcogenides is their nonvolatility, such that their properties can tune and hold their states in a self-holding manner. Thus, the chalcogenide glass materials enable active reconfiguration of optoelectronic memristors and photonic synaptic neural networks. For the latter application, a new class of transparent chalcogenide phase change material (PCM) exemplified by $Sb_2Se_3$, compatible with silicon photonic foundry process integration, will be explored. $Sb_2Se_3$ offers a large index contrast ($\Delta n$ = 0.8) and an extinction coefficient ($1.4 \times 10^{-4}$) orders of magnitude smaller compared to existing PCMs (e.g. GST, $VO_2$). We have demonstrated the first electrically driven photonic switch based on $Sb_2Se_3$ with a negligible excess loss of 0.0055 dB/$\pi$, implying its applicability to large-scale photonic synaptic network arrays (**Fig. 6ab**). Experimental and theoretical studies of the in PCMs, ferroelectric, ferromagnetic, semiconducting oxides, and ionic graphene materials [42]–[45] at molecular levels reveal molecular behaviors similar to those seen in biological synapses.

### B.2.2. Opto-electronic Neurons & Interconnects:

The construction of the optoelectronic neurons in the photonic neural network (PNN) includes photonic-memristive dendrites, photonic-memristive synapses, photonic axons, and nano-electronic SOMAs. Recent work successfully demonstrated bio-plausible optoelectronic neurons that match the Izhikevich's neuron model [38], [41] by simulation [53]–[56] and by experiment [39]–[41], [57], [58]. As **Fig. 7 (c-d)** shows, simulation results using HSpice in **Fig. 7 (c)** match the experimental results in **Fig. 7 (d)** obtained from prototyped optoelectronic neurons fabricated utilizing off-the-shelf components and from commercial foundry processes [57], [58]. Ultimate energy efficiency can be obtained from nanoscale optoelectronic neurons (**Fig. 7 (a)**) consisting of nanophotonic detectors, nanolasers [59], and nanoFETs with extremely low parasitics (< 0.5 fF) [60] for < 1fJ/spike efficiency (~10 fJ/spike for > 80× fanout), integrated on a silicon photonic platform exploiting quantum impedance conversion [61] capable of exhibiting both excitatory and inhibitory neuron behaviors by using nanophotonic detectors PD1 and PD2, respectively. The optoelectronic circuits can further integrate K+, Na+, and Ca++ ion-channels consistent with the neuroscience data.

As **Fig. 8** illustrates, these OE neurons are interconnected with each other in a photonic synaptic network consisting of photonic memristive-MZI synapses shown in **Fig. 6** to create a photonic neural network (PNN). Future 3D integration of photonic networks can benefit from optical TSVs (TSOVs) utilizing WDM for high fan-in/fanout interconnections mimicking the brain. For monolithic 3D EPIC, the recent work [62] demonstrated silicon photonic TSOVs exploiting 45° anisotropic etching of silicon (**Fig. 9**). For photonic interposer-based 3D integration of 3D-EPICs, TSOVs achieve less than 1-dB loss across optical bandwidths beyond 300 nm, enabling multi-wavelength (WDM) 3D photonic interconnection between the EPIC planes and between 3D EPIC chips. In other words, each of such photonic TSOVs can carry hundreds or even thousands of wavelengths to support and enhance the fanout of the photonic neural networks as well as the underlying electronic neural networks. An optoelectronic *Ising machine* can be realized on a PNN by utilizing a PTNO as a soma integrated with a nanolaser (E/O) and nanodetectors (O/E) to complete a parametrically-driven optical oscillator neuron (**Fig. 10b**). The working principle of the emergent dynamics will be the same as discussed in the previous section and allows a scale-free hierarchical photonic-electronic CTDS based on collective dynamics of OE devices and their coupled networks to achieve distributed learning.

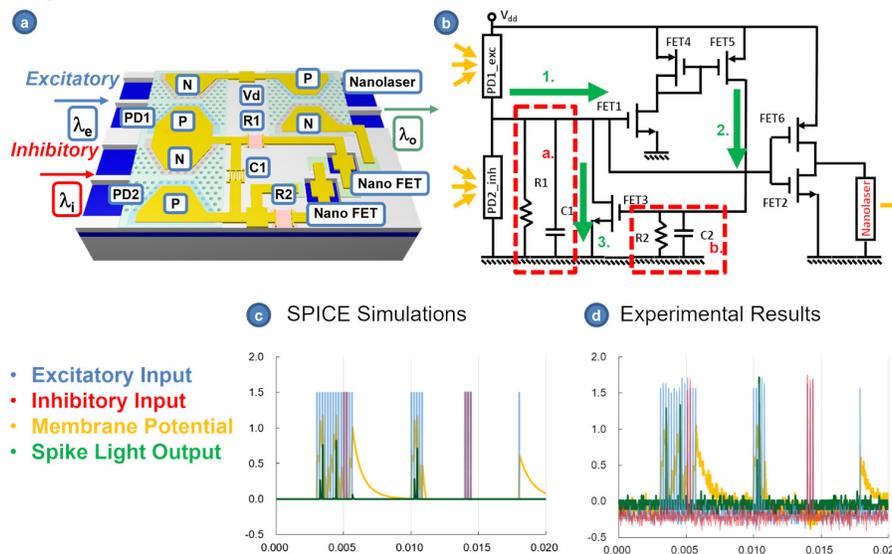

**FIGURE 7.** [41], [63], [64] (a) Nanoscale optoelectronic spiking neurons consisting of nanoscale nanophotonic detectors, nanolasers, and nanoelectronics with extremely low parasitics (<0.5fF) for <1fJ/spike efficiency integrated on a silicon-photonic platform. (b) spiking optoelectronic nano neuron circuit design using Cadence circuit-level simulation using the Verilog-A model, (c) Neuron spiking behavior with both excitatory and inhibitory signal inputs showing HSpice simulation result and (d) experimental results of neuron behavior with optical I/O.

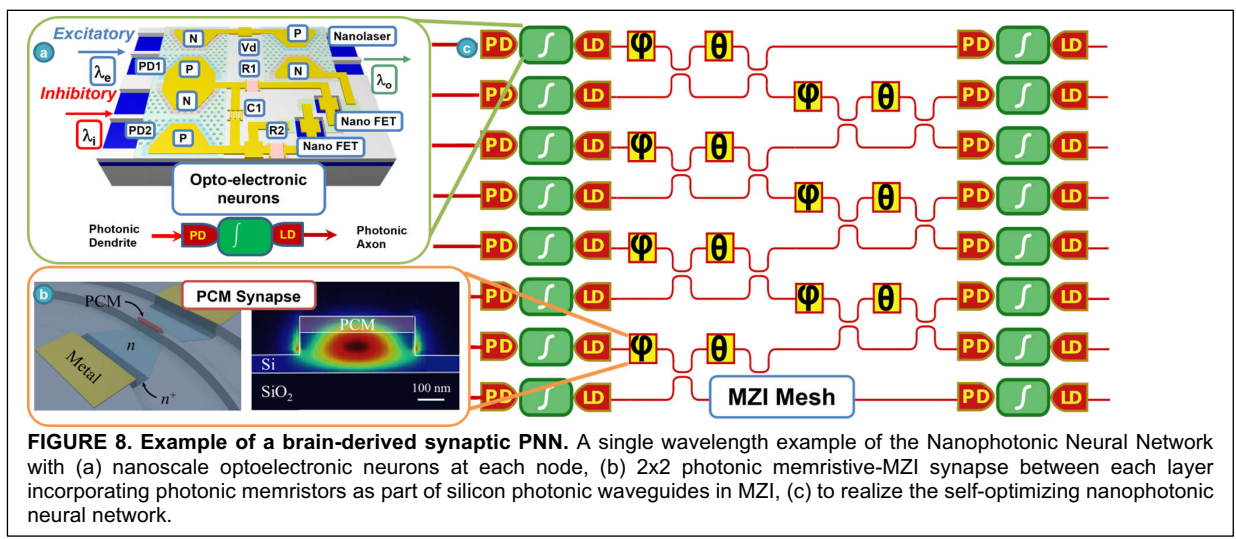

**FIGURE 8. Example of a brain-derived synaptic PNN.** A single wavelength example of the Nanophotonic Neural Network with (a) nanoscale optoelectronic neurons at each node, (b) 2x2 photonic memristive-MZI synapse between each layer incorporating photonic memristors as part of silicon photonic waveguides in MZI, (c) to realize the self-optimizing nanophotonic neural network.

### B.3. 3D Integrated Neuromorphic Computing Circuits and Architecture

***Background:*** Neuromorphic computing requires extremely high scalability and large connectivity while minimizing energy consumption and size. Human brains are structured in 3D and consist of multiple regions interconnected with each other to facilitate hierarchical learning. Previous attempts of neuromorphic computing research failed to reverse-engineer the brain due to a lack of scalable and energy-efficient interconnecting circuits for brain-like structured hierarchical learning (Gap 4 of **Table 1**). 3D integration of electronic and photonic integrated circuits offers high density and high connectivity with extreme efficiency at scale while supporting hierarchical learning in optical macro-circuits and electronic micro-circuits.

***Objectives:*** We should integrate electronic neural networks, photonic neural networks, and learning algorithms into a 3D EPIC neural network to achieve human-like hierarchical learning capability noting the following:

- Electronic neural networks and photonic neural networks can work independently, in parallel, or in hierarchy. The optoelectronic neurons are essentially electronic neurons with photonic/electronic dendrite terminals (photodetectors) and electronic/photonic axon terminals (lasers or modulators), hence, the electronic and photonic neural networks can work together or separately depending on the electronic synaptic connections between the OE neurons and electronic neurons. Hence, the learning process (e.g. Hebbian learning) can take place across the electronic and photonic neural networks.
- We can build 3D EPICs based on brain-derived principles to support scalable hierarchical learning networks with self-learning (e.g., predictive-error learning), resembling the human brain.
- 3D EPIC with photonic tensor cores can also adopt sparsity of the human brain to further enhance scaling and energy efficiency; human brain's energy efficiency and scalability are enhanced by pruning of synaptic interconnects through iterative predictive-learning processes and empirical experiences.

3D EPIC offers a new platform that combines *low-noise, scalability, wavelength-parallelism, high-throughput, and dynamic memristive plasticity* of **photonics**, and the *high-density integration, agility, and dynamic plasticity* of **electronics.** While the electronic neuromorphic computing approaches alone are unable to achieve the vast connectivity (e.g. ~8000 synaptic connections per neuron), at scale (e.g. billions of neurons), at the energy efficiency of the human brain and while the photonic neuromorphic computing approaches alone are unable to achieve the high density (100 billion neurons in 3 liter volume), *the 3D EPIC* is expected to achieve brain-derived neuromorphic computing with performance, efficiency, and density approaching those of a human brain.

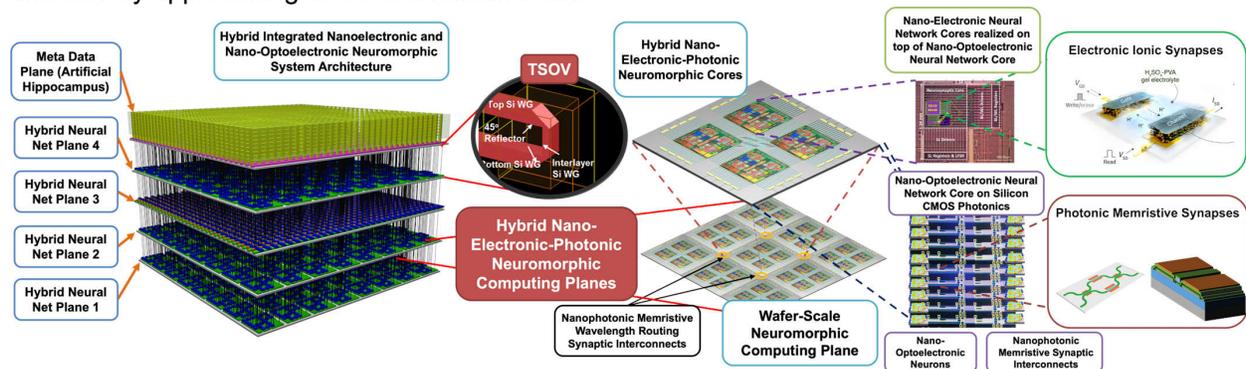

**FIGURE 9.** Schematic of the conceptual hybrid neuromorphic computing platform consisting of the wafer-scale nano-optoelectronic neural network intimately integrated with heterogeneous nano-electronic neural networks on top (shown are Stanford/UCSD electronic neural network die [65] examples). Each of the nano-electronic neural networks are in each section (~5 mm x 5 mm) on the wafer-scale nano-optoelectronic neuromorphic processor. Each section is interconnected to each other via $M \times N$ nanophotonic synaptic interconnects ($M$, $N$~1000). Each wafer will be 3D stacked utilizing 3D optical vias (TSOVs) described in [62].

### B.3.1. 3D EPIC Hardware Design, Fabrication, and Integration:

The conceptual brain-derived neuromorphic system (**Fig. 2**) consists of the hybrid neuromorphic computing platform shown in **Fig. 9** stacked in 3D hierarchy. Each plane consists of the wafer-scale nano-optoelectronic neuromorphic macro-circuits intimately integrated with heterogeneous nano-electronic neuromorphic micro-circuits integrated vertically by the Back-End-Of-The-Line (BEOL) processes described in **Fig. 5 &10c**. Each electronic plane links to each section of the nano-optoelectronic macro-circuit, and multiple such sections are interconnected to each other in parallel wavelengths via $M \times N$ photonic wavelength routing WDM synaptic interconnects ($M, N$ >1000 [66]). The resulting hybrid neural network planes are vertically stacked in 3D utilizing TSOVs [62], which can support transport and routing of photonic signals on 100s of wavelengths. Such wavelength routing capabilities and 3D photonic-electronic hierarchical circuits enable the neuromorphic computing networked system depicted in **Fig. 2k &10a** to achieve hierarchical learning.

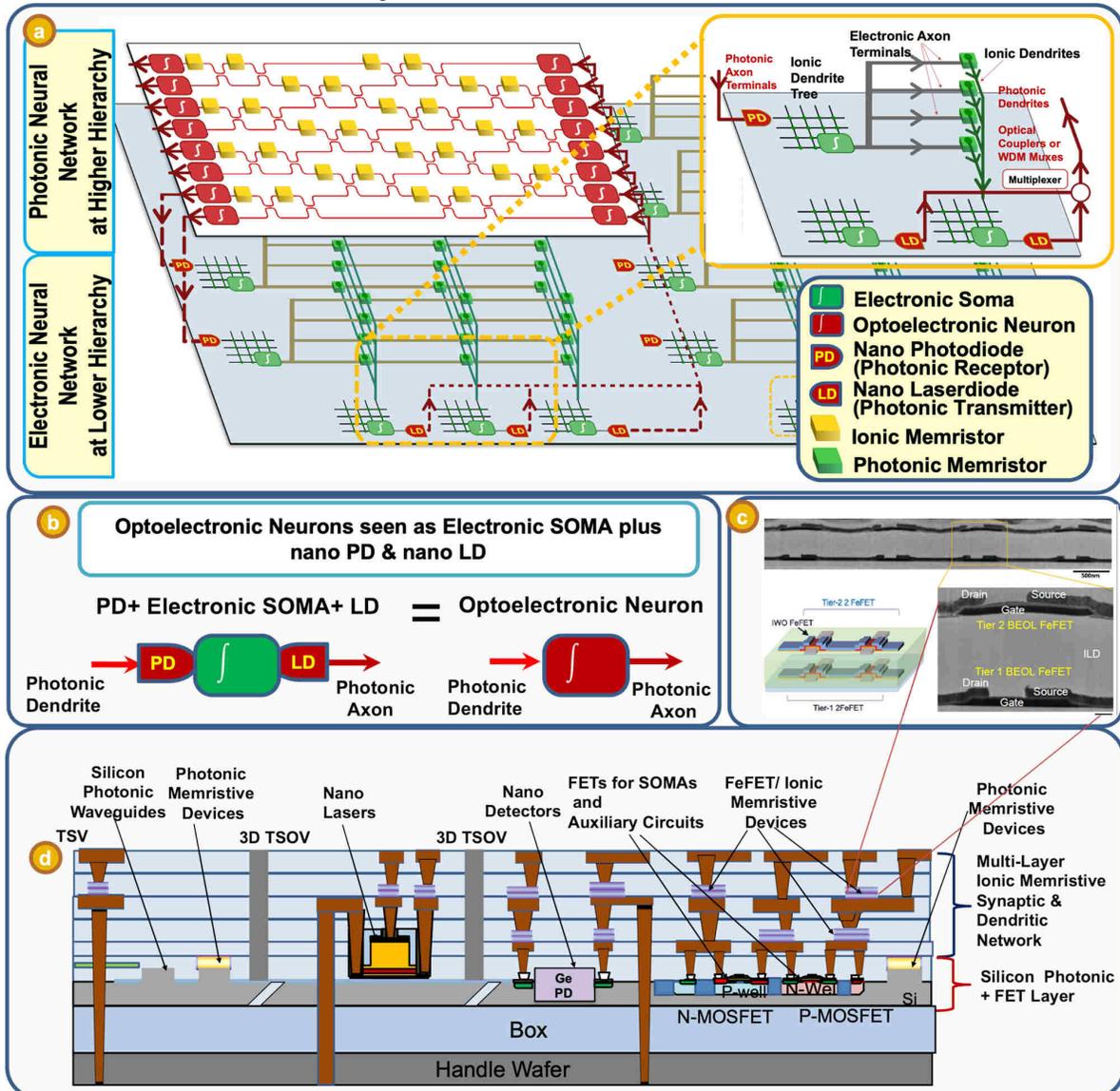

**FIGURE 10.** (a) Hierarchical photonic and electronic neural networks consisting of ionic dendrites, electronic somas, electronic axons, and ionic synapses for the electronic neural networks, and photonic memristive synapses, photonic dendrites, photonic somas, and photonic axons for photonic neural networks [80] Some of the electronic somas are equipped with nanoscale photodetectors and lasers to interface with photonic synapses at the higher hierarchy. (b) optoelectronic neurons consist of nanoscale photodetectors (excitatory and inhibitory), electronic somas, and nanoscale lasers, (c) a single plane of the hierarchical photonic and electronic neural networks is based on the silicon photonic + FET layer for photonic neural networks, electronic somas, auxiliary circuits, and ionic memristive materials and ECRAMs on BEOL metal layers. The electronic through-silicon-vias and photonic through-silicon-optical-vias (TSOVs) allow us to stack this plane in multiple stacks to complete a 3D EPIC [1][69]. (d) BEOL FeFET that can create electronic 3D neural network described in Thrust 2, where Datta successfully created 3D stacks of FeFET synaptic networks.

The new approach discussed in this paper will pursue a brain-derived, hierarchical, and reconfigurable neuromorphic computing system architecture co-designed with intelligent materials, devices, and learning algorithms. The new brain-derived neuromorphic system consists of the hybrid neuromorphic computing platform shown in **Fig. 9** in 3D according to the following biological principles: (1) Each electronic crossbar implements the learnable synapses for one projection within the hierarchical network architecture (e.g. **Fig. 3 and Fig 8a**), with *N* sending neurons and *M* receiving dendritic branches; (2) The conductance of each dendritic branch per neuron results in a current that integrated by the soma according to configurable nonlinearities, to be determined by the algorithm research, the net current contribution to the soma is integrated over time to produce the somatic membrane potential, $V_m$; (3) The somatic $V_m$ value then drives discrete spiking via excitable dynamics with a floating threshold that initiates a broadcast of the spike to all relevant electronic crossbar circuits; (4) In parallel, the optical interconnect broadcasts shortcut connections widely across the PNN using high-bandwidth WDM or optical couplers. Following these principles **Fig. 10a** shows a logical hierarchical network of photonic and electronic neural networks shown in two different hierarchical planes, although they are physically in the same hybrid plane in **Fig. 10c**. As **Fig. 2M** indicates [15], the photonic neural networks are inspired by the long-range connections between cortical regions shown from the macaque brain [16]; while the neurons are oblivious to the two different physical neural networks, photonic neural networks will achieve long-range communications between the remote neurons at far lower energy, latency, and noise. The electronic neurons include electronic (e.g. ionic polymer) dendrites with or without nano-photodetector receptors (excitatory or inhibitory), electronic axons with axon terminals with or without nanolasers, so that electronic somas can interface either with multiple planes of the ionic memristive synaptic interconnection networks or with photonic synaptic interconnection networks. The interface to the photonic synaptic interconnection networks includes photonic dendrites that are optical couplers to the photonic transmitters (nano-lasers) with wavelength multiplexers or simple optical power couplers. The ionic dendrite trees achieve dendrite computing, essential for low-energy, high-density, and scalable computing [70]. The OE neuron (**Fig. 10b**) consists of electronic soma enhanced with a dendrite with a nano-photodetector receptor (excitatory and/or inhibitory) and a transmitter with nanolaser at the photonic axon. **Fig. 7ab** shows an optoelectronic neuron example with a simple leaky-integrate-and-fire (LIF) electronic soma. The photonic synaptic interconnection networks and photonic neurons constitute photonic spiking neural networks. The photonic-electronic hierarchical neural network achieves symbiotic and synergistic neuromorphic computing across photonic and electronic neural networks (**Fig. 10a**).

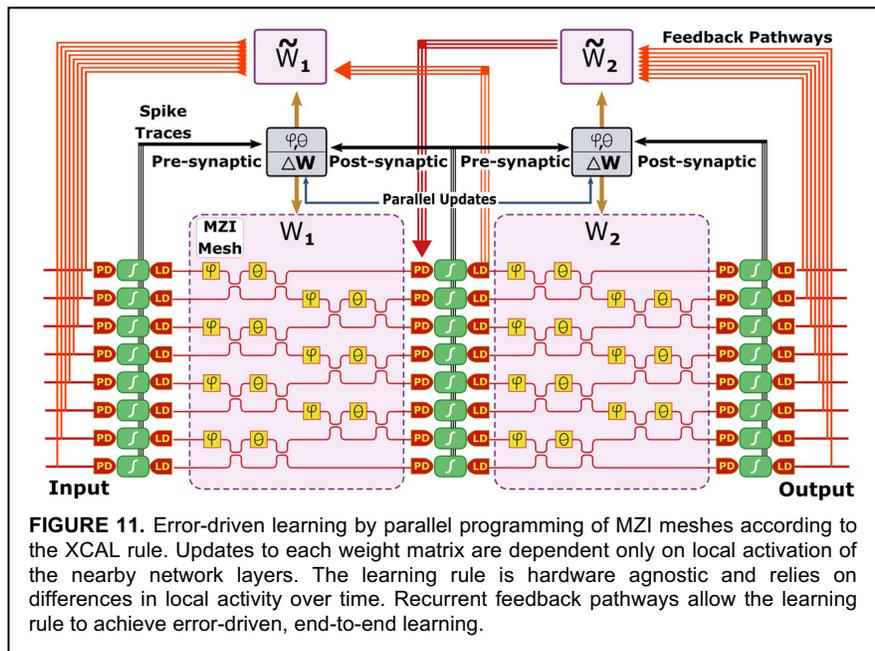

**FIGURE 11.** Error-driven learning by parallel programming of MZI meshes according to the XCAL rule. Updates to each weight matrix are dependent only on local activation of the nearby network layers. The learning rule is hardware agnostic and relies on differences in local activity over time. Recurrent feedback pathways allow the learning rule to achieve error-driven, end-to-end learning.

As exemplified in **Fig. 7**, nonlinear and linear dynamics of all elements can utilize Verilog-A and HSPICE modeling for co-design of photonic, electronic, and ionic components. In particular, interfacing between nanotransistors (FETs) and photonics can be realized by building the photonic Verilog-A models and importing them into electrical circuits on HSPICE simulations. This optoelectronics-based modeling technique can bridge the OE conversion and simulate soma and dendrites' space and time information. In the electronic neural networks, we will include models based on advanced low-noise FETs [71] with full

Process Design Kits (PDKs) capable of conducting HSPICE modeling for abstraction for large-scale circuit modeling [41].

Although photonic neural networks and electronic neural networks can operate independently, electronic-only neural networks cannot scale without excessive noise and power-penalty by themselves, while photonic-only neural networks cannot achieve high density or include time-delays easily. As **Fig. 10d** illustrates, each plane of 3D EPIC consists of the wafer-scale nano-optoelectronic neuromorphic macro-circuits intimately integrated with heterogeneous nano-electronic neuromorphic micro-circuits. The platform already contains the auxiliary silicon CMOS circuits and silicon photonics incorporating dynamic photonic materials, and BEOL post-fabrication on top of the metal layer will allow integration of multiple layers of electronic neural networks consisting of dynamic electronic/ionic materials and devices (Task 2.3). **Fig. 10c** shows BEOL FeFET fabricated by Datta for 3D electronic FeFET synaptic networks that can be formed on the first or second metal layer (M1 or M2) of silicon CMOS-Photonic neural networks. PI Yoo recently demonstrated a CMOS-photonic neural network that contains CMOS auxiliary circuits and silicon photonics [63], [72] Each wafer can be 3D stacked utilizing 3D TSOVs described in [73], [74]. 3D TSOVs achieve low loss vertical coupling between photonic layers to complete 3D EPIC hybrid neural networks of **Fig. 9**.

### *B.3.2. Learning algorithm integration with 3D EPIC: On-Line, Self-Supervised Learning in hierarchical 3D EPIC neural networks:*

To train these hybrid neural networks for specific tasks, we need training algorithms that can handle deeply recurrent neural networks with spiking neurons. Bio-plausible neuromorphic learning algorithms require only information that is local to a given synaptic connection, and this locality is important in the context of recurrent neural networks that would otherwise be "unrolled" over time to employ traditional backpropagation-based algorithms. A variety of such local learning rules exist that are capable of unsupervised learning based on temporal associations—for example, the simplest spike-timing-dependent plasticity rules (STDP). However, such unsupervised learning methods have not shown much success compared to traditional deep learning methods on various tasks, and it is clear that some form of error-driven supervised training scheme is necessary.

Extended Contrastive Attractor Learning (XCAL) and other functionally equivalent learning rules are proven equivalent to error-driven learning in recurrent neural networks [75]. The XCAL rule is technology-agnostic and allows end-to-end training of a neural network using only the activity of the sending and receiving neuron for each weight [76], therefore this learning rule is both local and error-driven:

$$\Delta w = \begin{matrix} (xy - \theta_p) & if \ xy > \theta_p \theta_d \\ -xy(1 - \theta_d)/\theta_d & otherwise \end{matrix}$$

where $\theta_p$ is a floating threshold parameter, $\theta_d$ is a constant, and $xy$ represents the product of moving time averages of activity for the sending and receiving neurons, respectively. In this scheme, all weight updates can be calculated in parallel without the complexity and explicit knowledge of any gradients within the network. The learning rule is also tolerant to differences in the temporal dynamics of the pre- and post-synaptic neurons and can be temporally scaled according to the temporal structure of the task to be completed.

As **Fig. 11** illustrates, a neural network can be trained in a supervised manner using a single feedforward and feedback pathway between each layer. A training sample is first fed to the network from the input layer and the network activity is allowed to stabilize. Next, the desired output activity is forced onto the output layer and network activity settles into a new pattern. Differences in activity over time pair with the XCAL learning rule to drive error-driven learning. In human brains, the observation of causal outcomes in the environment follows this same general pattern and allows for predictive, error-driven learning [75], [77]—in other words, the brain generates its own ground truth and is therefore self-supervised. Typical interactions with the environment occur much more slowly than the reconfiguration speed of optoelectronics. As such, weight updates can be calculated in an abstract manner and then quickly mapped to the implemented hardware to form intelligent devices. In **Fig. 11**, electrical signals carry spike timing information from optoelectronic neurons to an XCAL arithmetic control unit, which maps weight updates onto the phase shifters for each Mach-Zehnder interferometer of the local mesh.

### B.4. Learning System Integration and Experimental Testbed

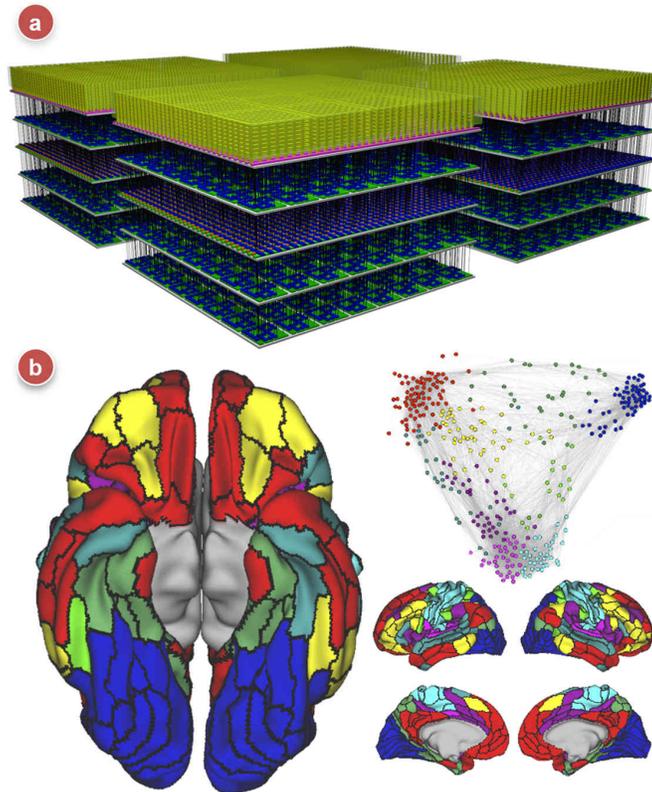

**FIGURE 12. Multiples of 3DEPICs in functional blocks. (a)** Multiples of the 3D EPICs can be interconnected and can be structured to adopt the neuroscience architectures (structured brain circuits) and algorithms developed for self-supervised learning (e.g., predictive-error learning). **(b)** Force-directed graph showing intrinsic connectivity of cortico-hippocampal networks in the human brain [78].

***Background:*** To date, neuromorphic computing platforms have been tested against specific benchmark tasks in isolation—for example, handwriting recognition. The human brain, in contrast, handles multiple tasks, and with a level of anatomical complexity that is orders of magnitude beyond prior implementations of neuromorphic computing. Brain-derived neuromorphic computing, implemented at a large scale, could potentially enable more complex, intelligent behavior, and emergent properties in such a system could provide new insights into the relationship between brain structure, physiology, and behavior.

***Objective:*** We should use basic neuroscience to inform the development of neuromorphic computing hardware that will be implemented at a scale that is sufficient to handle complex tasks. We will create a prototype neuromorphic computing hardware system that incorporates brain-derived algorithms and structure to perform complex tasks thought to require human-level intelligence. The explicit mechanisms in the neuromorphic hardware will then be subjected to detailed measurements and highly targeted causal manipulations to generate new theories of how brain function emerges from interactions between cells and network systems over time. The iterative process will reveal new insights into how variations in system specification affect representations, dynamics, communication, and cognitive capabilities as follows:

- The brain-derived hardware utilizing Multiple 3D EPICs integrated with brain-derived algorithms will be the first of its kind. It will not only simulate the brain in structure and function, but will also serve as a prototype brain that can be causally manipulated to examine mechanisms of learning derived from the biological brain.
- The use of an identical experimental testbed (3D VR navigation) for both the neuroscience studies and for the 3D EPIC prototype will enable iterative development of neuro-morphic computing technology and mechanistic models of human brain function. The prototype will enable us to test whether the assumptions derived from neuroscience suffice to give rise to intelligent behavior. Measurements and manipulations in the prototypes, in turn, can reveal emergent properties that could lead to transformative advances in basic neuroscience, addressing questions regarding the neural code and the links between cellular, circuit, network, and behavioral levels of analysis. Even development failures in the prototype will illuminate the capabilities and limitations of a scaled-up, brain-like neuromorphic computing system, and provide insights into biological mechanisms that give rise to behavior (*Gap 1*).

We envision an iterative process, by which neuroscience research guides development of the neuromorphic computing architecture, and measurements and manipulations in the prototype address questions and controversies in basic neuroscience.

### *B.4.1. A Brain-Derived Neuromorphic Computing System capable of fast learning, generalization, and problem solving:*

First, we will integrate learning algorithms from the large-scale computational model of navigation into the Brain-Derived Prototype systems consisting of **multiple** 3D EPICs (**Fig. 12a**). Circuits will be structured to emulate the neural architecture of hippocampus, thalamus, prefrontal cortex, and other sites (**Fig. 12b**). The 3D EPICs in **Fig. 12a** will be interconnected and trained to be functionally homologous to the local and large-scale networks of the brain. Then, we will further integrate multi-modal sensory functions (visual, audio, etc.) and motor functions (Unmaned Aerial Vehicles-UAVs) with the multiple 3D EPICs to create a prototype system.

The resulting prototype system will be trained and probed experimentally for evidence of hierarchical and self-supervised learning (e.g., predictive error-driven learning) mechanisms that are present in data and the large-scale computational model developed by O'Reilly et al. Test of the prototype will occur first using simulation testbeds first, followed by physical navigation in real-world environments that require sensory and motor (effector) systems. These experiments will enable us to examine the performance of the algorithms implemented in the computing hardware with pixel level sensory inputs that change with actions implemented by the device. Benchmark tasks will start simple and gradually increase in complexity; they will consist of the following:

(1) <u>Elementary perception and movement.</u> The prototype system will first undergo basic training and inference testing until it demonstrates accurately controlled movement direction, distance, and speed in the virtual environment analogous to biological systems. Success on this task will require development of sensory-motor systems capable of the following: visual sensations leading to object recognition, selective attention to relevant objects for action, learning and remembering explored environments, and movement through space-based goals mapped to knowledge. <u>(2) Basic navigation</u>. The prototype will be trained on complex behaviors evident in all the model biological systems (rats, humans). (a) Beaconing and Foraging: efficient sampling of visual information for locating navigational landmarks and planning efficient movements to retrieve multiple targets and reach a goal location. (b) Route-based navigation: Reaching a goal based on instructions that specify specific actions at particular landmarks. (c) Map-based navigation: Learning to navigate to an unmarked goal location from varying starting points. Benchmark tasks for each of these three behaviors will increase in complexity (# of steps) in order to discover boundary conditions of system capabilities. (3) <u>High level cognition</u>. (a) Compositionality: Leverage prior learning and recombine elements from previously learned navigation tasks into new sequences of goal-directed behavior. (b) Inference: Spontaneously draw upon prior learning of the environment to infer new optimized routes to goal locations. (c) Flexibility: When confronted by novel barriers, efficiently sample relevant information in order to spontaneously plan and execute alternative routes. (d) Generalization: Leverage prior knowledge to rapidly learn the structure of a new environment, task, and goal by optimally applying a subset of existing information to current demands of new tasks. (e) Abstraction: Use hierarchical goal representations to realize specific goals and subgoals (e.g., finding a sandwich) and to realize an abstract goal (e.g., find something to eat).

### *B.4.2. Understanding the brain with Neuromorphic Prototype Systems:*

Once we create such a neuromorphic prototype system (or microsystem), we can: (1) identify properties that emerge from the fully connected brain-derived computing system, and compare these properties to comparable measurements taken in experiments, and then (2) perform controlled experiments with causal manipulations that could not be done with human or rodent subjects, in order to relate specific mechanisms to observable behavior.  The prototype system allows *Causal Manipulations* of (1) independent excitation of ion channels (Na+, K+, and Ca2+) of any synapse at any timescale, (2) the timing and threshold of firing in each neuron, (3) the plasticity, strengths, and dynamics of each synapse, (4) the topology of neural circuits, (5) the speed of spike propagation (e.g. myelinated vs unmyelinated axons for photonic vs. electrical wires), etc.. The prototype system also enables *Measurements* of (1) membrane potentials of every soma, (2) spike outputs at every axon, (3) synaptic strengths and their dynamic behavior, (4) propagations of spikes across the neural network, (5) forensic tomography of spikes and energy consumption in neural networks, etc.

These unique capabilities in the prototype systems will enable us to understand links between specific neuronal and network level properties with system-level functioning, and behavior. Among the many possible questions that can be addressed, three specific examples include: (1) <u>Learning processes in sparse hierarchical networks.</u> As described earlier, the 3D EPIC utilizes hierarchical electronic and photonic networks with tensor-core decompositions[79] . We can manipulate the network connectivity characteristics in a manner that mimics the developmental trajectory of cortical networks, and we can determine how alterations in this trajectory affect "neural" measurements, learning, and behavior in the prototype. (2) <u>Oscillations and long-range communication and coordination.</u> Theta, alpha, beta, and gamma oscillations are prevalent in many neurophysiological brain recordings, but there is intense controversy about whether these rhythmic changes in population-level excitability play a necessary role in plasticity, network-level interactions, and cognition. We can investigate how oscillations emerge in the neuromorphic system and compare these results to results from our investigations. We can then conduct causal manipulations to neuronal characteristics and network-level connectivity in order to identify factors that give rise to different kinds of oscillations and determine how these oscillations affect learning and behavior in the prototype. (3) <u>Brain Network Efficiency</u>: we can benchmark learning capabilities and energy-efficiency of neural network prototypes of various topologies and morphologies. Pruning of synaptic connections, controlling of synaptic plasticity, or introducing more diversity in neurons in the prototypes in this study may lead to an optimization or a tradeoff between learning performance and energy-efficiency. These studies can reveal how the brain is capable of accomplishing massive computational tasks with high energy efficiency.

## C. Summary

Reverse-engineering the brain, if successful in creating a brain-derived neuromorphic computing system, can lead us to profound changes in the way we lead daily lives not only by providing flexible and adaptive learning to assist us but also by helping us understand how the brain works. While the past efforts of analog and digital electronics approaches to reverse-engineering the brain have not resulted in realizing brain-like computing, a new brain-derived neuromorphic computing approach involving bio-realistic dynamicity at every level--- atom/molecular, neuronal/synaptic, microcircuit, and multi-structural microcircuit levels --- can possibly realize reverse-engineering the brain. New generations of nanoscale memristive ionic, electronic, and photonic materials can possibly resemble the dynamic mechanisms responsible in the biological neural systems. Utilizing those materials, we can pursue new photonic, electronic, and ionic memristive devices that can closely resemble the dynamic mechanisms seen in the biological neural circuits and systems. We can then pursue 3D photonic-electronic integrated circuits that offer high density and high connectivity with extreme efficiency at scale while supporting hierarchical learning in optical macro-circuits and electronic micro-circuits. A neuromorphic computing simulator developed using these new materials, devices, and circuits based on the bio-plausible principles can help us design the prototootype systems and conduct experimental testbed studies involving them. The unique capabilities in the prototype systems will enable us to understand links between specific neuronal and network level properties with system-level functioning, and behavior. Among the many possible questions that can be addressed, three specific examples include: (1) <u>Learning processes in sparse hierarchical networks.</u> (2) <u>Oscillations and long-range communication and coordination.</u>, and (3) <u>Brain Network Efficiency</u>. These studies can reveal how the brain is capable of accomplishing massive computational tasks with high energy efficiency.

## D. Acknowledgements

This material is based upon work supported by the Air Force Office of Scientific Research under award number FA9550-18-1-0186 and award number FA 9550-22-1-0532, and in part by the Office of the Director of National Intelligence, Intelligence Advanced Research Projects Activity under Grant 2021-21090200004. The authors acknowledge enlightening discussions with Prof. H.-S. Philip Wong of Stanford University.

Waveguide Router Interposers and Silicon Photonic Transceivers," *IEEE Journal of Selected Topics in Quantum Electronics*, vol. 25, no. 5, Sep. 2019, doi: 10.1109/JSTQE.2019.2910415.